\documentclass[preprint,showpacs]{revtex4-1}
\date{November 23,  2011}  
\usepackage{amsthm}
\usepackage{amsmath}
\usepackage{amssymb}
\usepackage{color} 

\newtheorem{theorem}{Theorem}
\newtheorem{lem}[theorem]{Lemma}
\newtheorem{cor}[theorem]{Corollary}

\newtheorem{proposition}[theorem]{Proposition}

\numberwithin{equation}{section} 
\numberwithin{theorem}{section}
\renewcommand{\theequation}{\arabic{section}.\arabic{equation}}

\def\Zd{\ensuremath{\mathbb{Z}^d}}
\def\Fields{\ensuremath{\mathcal{E}}}
\def\R{\ensuremath{\mathbb{R}}}
\newcommand{\N}{\ensuremath{\mathbb{N}}}
\DeclareMathOperator{\Tr}{Tr}
\DeclareMathOperator{\diam}{diam}

\newcommand{\gf}[2]{\ensuremath{\widehat{G}_{#1}^{#2}}}
\newcommand{\lop}{\ensuremath{\kappa}}
\newcommand{\F}{\ensuremath{\mathcal{F}}}
\newcommand{\ham}{\ensuremath{\mathcal{H}}}

\def\bc{B}

\def\alg{\mathcal{A}}

\DeclareMathOperator{\Av}{Av}
\newcommand{\lip}[1]{\ensuremath{\left| \left| \left| #1 \right| \right| \right|}}
\newcommand{\condAv}[2]{\ensuremath{\Av\left[ #1 \middle| #2 \right]}}
\newcommand{\pnorm}[2]{\ensuremath{\left\| #1 \right\|_{#2}}}

\newcommand{\overlappingClass}[2]{#1 \cap #2 \notin \{ \emptyset,#1 \}}
\newcommand{\vecEtaCond}[1]{\ensuremath{\hat{e} \cdot \vec{\eta}_{#1}}}

\newcommand{\piecewise}[1]{\ensuremath{ \left\{ \begin{array}{ll} #1 \end{array} \right.}}
\newcommand{\state}[1]{\ensuremath{\left\langle #1 \right\rangle }}

\newcommand{\Doi}{} 

\DeclareMathOperator*{\LIM}{LIM} 

 \def\Z{\mathbb Z}
 \def\be{\begin{equation} } 
  \def\ee{\end{equation} }
\def\1st{$1^{\textrm{st}}$} 

\begin{document}
\title{Proof of Rounding by Quenched Disorder of First Order Transitions in Low-Dimensional Quantum Systems}
\author{Michael Aizenman}
\email{aizenman@princeton.edu}
\affiliation{Departments of Physics and Mathematics, Princeton University, Princeton NJ 08544-8019}
\author{Rafael L. Greenblatt}
\email{greenbla@mat.uniroma3.it}
\affiliation{Dipartamento di Matematica, Universit{\`a} degli Studi Roma Tre, Largo San Leonardo Murialdo 1, 00146 Roma, Italy}
\author{Joel L. Lebowitz}
\email{lebowitz@math.rutgers.edu}
\affiliation{Departments of Mathematics and Physics, Rutgers University, Piscataway NJ 08854-8019}

\begin{abstract} 
We prove  that for quantum lattice systems 
in $d\le 2$ dimensions
 the addition of quenched disorder rounds any first order phase transition   
in the corresponding  conjugate order parameter, both at positive temperatures and at $T=0$.  
For systems with continuous symmetry the statement extends up to $d~\le~4$ dimensions.  This establishes for quantum systems the existence of the Imry-Ma phenomenon which for classical systems was proven by Aizenman and Wehr.  The extension of the proof to quantum systems is achieved by carrying out the analysis at the level of thermodynamic quantities rather than equilibrium states.
\end{abstract}
\maketitle

\section{Introduction}

Quenched disorder is known to have a pronounced effect on 
phase transitions in low dimensional systems.  As Imry and Ma pointed out in 1975~\cite{YM}, disorder can prevent the appearance of discontinuities, and of long range order, associated with first order transitions.   Nonetheless, for a number of years there was considerable uncertainty about the mechanisms involved in such a ``rounding effect'', and consequently also of the circumstances in which this effect appears~\cite{ParisiSourlas}.  Rigorous results in a series of works in the 1980s~\cite{Imbrie.PRL,*Imbrie.CMP,*BK.PRL,*BK.CMP,AW.PRL,*AW.CMP} established the conditions for rounding in classical systems.  The situation in quantum systems remained uncertain, with recent suggestions~\cite{Goswami} that for such the effect might not have the same general character as in the context of classical statistical mechanics.

In this work we present a modified version of the argument of Ref.~\onlinecite{AW.CMP} which allows us (as previously announced in Ref.~\onlinecite{QIMLetter})  to extend the classical results to quantum systems.  It was not obvious that this should be possible, since the statistical mechanical description of a $d$ dimensional quantum systems often has the appearance of a $(d+1)$ dimensional classical system, and in addition
some of the tools used in the classical case have no clear quantum generalizations.  The latter difficulty is resolved here by focusing on thermodynamic quantities, bypassing some of the subtle issues related to equilibrium states which played a role in Ref.~\onlinecite{AW.CMP}.  

Before presenting the formalism in which the general result is expressed, we start with some specific examples to which the results apply.   

\section{Examples of the rounding effect}\label{WhatIsRounding}

\subsection{Transverse Field Ising Model}\label{TFIM_sec}

The transverse-field Ising model in a random longitudinal field is defined by the Hamiltonian
\begin{equation}\label{TFIM_ham}
   \ham^{\textup{TFIM}} = -  J \sum_{\langle x,y \rangle}  \sigma_{3,x} \sigma_{3,y} - \lambda \sum \sigma_{1,x} - \sum (h + \epsilon \eta_x) \sigma_{3,x},
\end{equation}
where $\sigma_{j,x}$ are Pauli matrices describing the $j=1,2,3$  components of a spin at site $x$, and the first summation is over pairs of neighboring sites $x , y\in \Zd$.   The symbol $\eta_x$ represents a random field in the $j=3$ direction, whose values at different sites are given by independent and identically distributed (i.i.d.) random variables, satisfying certain mild regularity conditions (to be specified in Section~\ref{sec:main_results} below). 

In $d\ge 2$ dimensions the nonrandom version of the system (with $\epsilon=0$) exhibits  
spontaneous (or residual) magnetization 
 at $h=0$  if $J/\lambda$ and $J \beta$ are large enough.  In $d=1$ dimension this occurs only at zero temperature, $T\equiv \beta^{-1}=0$, i.e.\ in the ground state.  More explicitly, existence of \emph{spontaneous magnetization} means that 
 \be 
  \overline M(\beta) \ :=\ \frac{1}{2} [ M_{0+ }(\beta) \ - \    M_{0- }(\beta) ]\ > \ 0 
 \ee 
 where   $M_{0\pm}(\beta)$ are the 
two limits of the  Gibbs state expectation values in finite (rectangular) domains $\Gamma \subset \Z^d$, 
\begin{equation} \label{eq:M_pm}
  M_{0\pm}(\beta) := \lim_{h \to 0\pm} \lim_{\Gamma \nearrow \Zd} \frac{1}{|\Gamma|}\state{\sum_{x \in \Gamma}\sigma_{3,x}}^{h,\beta}_\Gamma
\, . 
\end{equation}
Since the $h$ limit is taken last, the boundary conditions on the finite systems do not affect the value of the limits in \eqref{eq:M_pm} (for the convenience of translation covariance our default choice is periodic boundary conditions). 
In view of the symmetry of the model, $M_{0+} \ = \ - M_{0-} \ = \   \overline M(\beta)$. 

The general result which is proven below (Theorem~\ref{alt_main_prop})  implies:
 
\noindent \emph{In $d \le 2$, the system described above, with the
Hamiltonian~\eqref{TFIM_ham},  at $\epsilon \neq 0$ has $ \overline M (\beta)=0 $ at all $\beta \le \infty$.  In particular, even if there is more than one ground state, they all yield the same bulk-average value for the mean magnetization.} 

This can also be restated in thermodynamic terms:   
Letting $\F(h,\beta)$ denote the thermodynamic limit of the specific free energy, 
which is defined below in \eqref{FDensDef}, 
the different magnetizations can be expressed as directional derivatives~\cite{Ruelle}
\begin{equation}
  M_{0\pm}(\beta) = \left. \frac{\partial \F}{\partial h \pm}\right|_{h=0} \, . 
\end{equation}
Non-vanishing spontaneous magnetization corresponds therefore to a discontinuity of a first order derivative of the free energy, here at $h=0$, and thus to a first order phase transition in the terminology of  Ehrenfest~\cite{Callen}.

Note that the order parameter   for the phase transition is the expectation of the volume average of the local operator $\sigma_{3,x}$ which in the Hamiltonian~\eqref{TFIM_ham} appears  coupled to the random field.  Such conjugacy is the essential condition for the general result  presented here.
The theorem does not imply that the rounding of the magnetization in the `$z$-direction' would occur when 
random terms are added only to the transverse field $\lambda$.  Indeed, as has long been known\cite{campanino1991lgs}, when only such disorder is present spontaneous magnetization does persist in two dimensions.   Such transversal disorder can have other subtle effects on the ferromagnetic-paramagnetic transition\cite{Fisher}, but these are beyond the scope of the present work.

\subsection{Isotropic Heisenberg model}

The isotropic, nearest-neighbor Heisenberg ferromagnet in a random magnetic field  is described by the Hamiltonian
\begin{equation} \label{HeisHam}
 \mathcal{H}^{\textup{Heis}} \ =\ - J \sum_{\langle x,y \rangle}  \vec{\sigma}_x \cdot \vec{\sigma}_y - \sum_x (\vec{h} + \epsilon \vec{\eta}_x) \cdot \vec{\sigma}_x
\end{equation}
where $J$ is a positive real number, $\vec{\sigma}_x$ is a vector spin operator associated with the site $x \in \Zd$, and $\vec{\eta}_x$ is a magnetic field which varies randomly from site to site.  We now assume that these random field variables are not only independently and identically distributed, but also that their distribution is rotation invariant.  We also assume that $\vec{\eta}_x$ is nonzero with probability~1.\\[1ex] 

In the case of a uniform magnetic field ($\epsilon=0$) it is universally believed that for $d \ge 3$ this system exhibits spontaneous magnetization at $\vec{h}=\vec{0}$ below some $J$-dependent critical temperature (although a rigorous proof is lacking).  Letting $\state{\cdot}^{\vec{h},\beta}_\Gamma$ now denote the Gibbs state of this system on a finite domain $\Gamma \subset \Zd$ with periodic boundary conditions at inverse temperature $\beta$, spontaneous magnetization can be expressed, similarly  to the previous cases, by the statement that  
\begin{equation}
  \vec{M}_0(\beta) := \lim_{h\searrow 0} \,  \lim_{\Gamma \nearrow \Zd} \frac{1}{|\Gamma|}\state{\sum_{x \in \Gamma}\vec{\sigma}_x}^{ h\hat{e},\beta}_\Gamma \ \neq \   0
\end{equation}
where $\hat{e}$ is any fixed unit vector (indicating the direction along which $\vec{h}$ is taken to $\vec{0}$).  
In terms of the thermodynamic limit of the specific free energy 
the  same quantity can be expressed as\cite{Ruelle}
\begin{equation}
  \vec{M}_0(\beta) \  =\   \lim_{h\searrow 0} \,  \,  \frac{d }{dh}    \F(h\hat{e},\beta) \, .
\end{equation}
Note that as in the previous example the randomness is conjugate to the quantity whose density is  the magnetization considered here.  

Theorem~\ref{alt_continuous_prop}, below, has the following implication for this system at $\epsilon \neq 0$:

\emph{
In $d \le 4$, the random-field Heisenberg ferromagnet described above has $\vec{M}_0(\beta)\ =\  0$ (and hence $\F(\vec{h},\beta)$ is differentiable in $\vec{h}$ at $\vec{h}=\vec{0}$) for any $\beta$ including $\beta = \infty$.   }
  
As will be seen in Section~\ref{Thermo_and_StatMech}, the vanishing of the spontaneous magnetization implies also the absence of ferromagnetic long range order. 

\section{A general formulation}\label{defs_section}

\subsection{The system and its Hamiltonian} \label{general_defs}
 
 We consider here systems on homogeneous $d$ dimensional lattices, which for simplicity we take to be $\Z^d$.  
Associated with each lattice site $x\in \Z^d$ is a quantum system whose state space is isomorphic to a common finite dimensional Hilbert space ${\mathcal H}_0$.   
 The local systems are coupled through a Hamiltonian which for a finite region $\Gamma \subset \Z^d $, with free boundary conditions, takes   the form
\begin{equation} \label{eq:H}
H_{\Gamma,0}^{h,\epsilon,\eta}=\sum_{A \subset \Gamma} Q_A + \sum_{x \in \Gamma} (h+\epsilon \eta_x)\lop_x  \, .  
\end{equation}
Here $Q_A$ is an operator which acts on the quantum degrees of freedom in $A$ (i.e. it is described by an operator acting in the space $\otimes_{x\in A} {\mathcal H}_x$).    The interaction is assumed to be  translation invariant ($Q_{T_x A} = U^\dagger_x Q_A  U_x$, with $T_x$ denoting translations on the lattice and $U_x$ the corresponding unitary operators).  
Unless stated otherwise it is  assumed here   
 that the interactions are of finite range (i.e., $Q_A=0$ for all sets with $\diam A > R$, at some finite $R$).   
However the stated results apply also to a  a broader class of long range interactions  with a sufficiently fast power law decay,  which is described  in Appendix~\ref{LR_appendix}.   
In particular,  Theorem~\ref{alt_main_prop}  applies to  two-body interactions satisfying $\|Q_{x,y}\| \le C /|x-y|^{3d/2}$, and Theorem~\ref{alt_continuous_prop} applies if $\|Q_{x,y}\| \le C /|x-y|^{\alpha}$ with some $\alpha > d-2$. 

Disorder is expressed in \eqref{eq:H} through the random coefficients  $\eta_x$.  Their  variance  is fixed at $1$, so that the strength of the disorder is controlled by the  parameter $\epsilon$.  
The terms $\eta_x$ multiply are assumed to be of the form   $\lop_x \equiv U^\dagger_x \lop_0  U_x $, with $\lop_0$ an operator which acts on a finite cluster of sites, whose size may be greater than $1$.
It should be noted that   $\{ \eta_x\} $ appear in \eqref{eq:H}  as additions to  a uniform parameter $h$.  This  parameter plays an essential role for the results presented here. 
The symbol $\eta$ without the subscript indicates the collection of random fields at all sites, and  $\eta_\Gamma$ denotes the random fields in some finite region $\Gamma \subset \Zd$.  
   In systems with several families of independent random terms our results hold for the observable associated with each family considered separately, but we will not discuss this point here (see \cite{RLG.thesis}).

Hamiltonians with other boundary conditions, $H_{\Gamma,\bc}^{h,\epsilon,\eta}$,  are defined analogously, so that the terms within $\Gamma$ are the same as for free boundary conditions, while at a finite distance from the boundary there can be additional  terms which are uniformly bounded in norm and in the size of the clusters of the directly affected sites.  Among the allowed options are periodic boundary conditions, denoted by$B=\textup{per}$, which  will be the default choice if no other is explicitly specified.

As usual, the finite volume partition function is denoted
\begin{equation} \label{Zdef}
Z_{\Gamma,\bc}(h,\beta,\epsilon;\eta) := \Tr \exp (- \beta H_{\Gamma,\bc}^{h,\epsilon\eta}),
\end{equation}
and the corresponding Gibbs state  is
\begin{equation} \label{stateDef}
\state{... }^{h,\beta,\epsilon}_{\Gamma,\bc} (\eta) := \frac{\Tr \left( ... \ e^{-\beta H_{\Gamma,\bc}^{h,\eta}}\right)}{Z_{\Gamma,\bc}(h,\beta,\epsilon;\eta) }  \, .
\end{equation}
When the values of the  superscripts is deemed obvious from the context,  they will sometimes be omitted.

\subsection{The quenched free energy and its  infinite volume limit} 

The free energy for a finite system is denoted here by 
\begin{equation}\label{FDef}
{F}_{\Gamma,\bc}(h,\beta,\epsilon;\eta) := \frac{-1}{\beta}  \log Z_{\Gamma}(h,\beta,\epsilon;\eta) \,  ,  
\end{equation} 
We will mostly be concerned with square domains $\Gamma_K = [-K,K]^d \cap \Z^d$ 
with the periodic boundary conditions, and so we introduce the abbreviation
\be \label{eq:per}
F^{h}_K (\eta)\  \equiv  \  {F}_{\Gamma_K,\textup{per}}(h,\beta,\epsilon;\eta)\, . 
\ee   
The free energy density, per unit volume, is
\begin{equation}\label{FDensDef}
{\mathcal F}_{\Gamma,\bc}(h,\beta,\epsilon;\eta) := F_{\Gamma,\bc}(h,\beta,\epsilon;\eta)  / |\Gamma|  \, .
\end{equation}
%
where $|\Gamma|$ is the number of sites in $\Gamma$.  As the notation suggests, for finite systems the  free energy depends on the choice of boundary conditions and on the disorder variables.  In the thermodynamic limit, however, the dependence of its density on the boundary conditions disappears and so does the dependence of its typical value on the disorder variables.  The following generally known result (we refer here to such statements as Propositions) applies to all systems of the type described above.  

\begin{proposition} \label{F_converges}
If the random fields are of finite variance and form a translation invariant and ergodic process, 
then for any  $\beta \in [0,\infty]$ there is a full measure set $\mathcal{N}$ of field configurations for which the infinite volume limit 
 \begin{equation}
\F(h,\beta,\epsilon) := \lim_{L \to \infty} {\mathcal F}_{\Gamma_L,\bc}(h,\beta,\epsilon;\eta)
\end{equation}
exists for all $h$ and its value is independent of $\eta$ and the boundary conditions $\bc$.
\end{proposition}

Since much of the early discussion of this result \cite{Ledrappier,*vuillermot1977tqr} was limited  to the classical case let us add that the argument used in that context extends directly also to quantum statistical mechanics.  In essence: the free energy density  can be approximated up to an arbitrarily small correction, of order $O(L_0^{-1})$ (see Inequality~\eqref{eq:FreeEnergyDifferences}) by  that of a system obtained by partitioning the space  into large blocks (of length $L_0$) among which the couplings were removed.  
For the approximants the density converges by either the law of large numbers, in case of independent $\{\eta_x\}$, or by the ergodic theorem in the more general case stated above.  That implies Proposition~\ref{F_converges}. 

The following remarks are of relevance:

{\it i.}   Since convexity is  automatically inherited by the limit, $\F(h,\beta,\epsilon)$ is concave as a function of $h$.

{\it ii.}  The case  $\beta = \infty$ corresponds to the ground state energy:  for almost every $\eta$ the limit 
\be \F(h,\infty,\epsilon) \  := \  \lim_{\beta \to \infty} \F(h,\beta,\epsilon)
\ee  exists and is equal to the limit of the energy densities of the finite volume ground states of the random Hamiltonian (whose value is independent of the boundary conditions).   In other words, for the free energy the limits $\beta \to \infty$ and $L\to \infty$ are interchangeable.  

{\it iii.}  The uniqueness of the free energy density  does not extend to uniqueness of the Gibbs states, or ground states in case $\beta = \infty$.    The question which our results address is whether the different Gibbs states (or ground states) of the given Hamiltonian can differ in their mean magnetization, that is in the volume averages of $\langle \lop_x \rangle$.

{\it iv.}  The following condition will allow us to weaken other assumptions.  We will say that our system satisfies the \emph{weak FKG condition with respect to $\lop$} if 
\begin{equation}
  \state{\lop_x}^h_\Gamma (\eta) \ge \state{\lop_x}^{h'}_\Gamma (\eta')
\end{equation}
whenever $x \in \Gamma$, $h \ge h'$, and $\eta_y \ge \eta'_y$ for all $y \in \Gamma$.  This condition is known to hold in the transverse field Ising model described in Section~\ref{TFIM_sec}  \cite{campanino1991lgs}.

\subsection{Quenched disorder with continuous symmetry} \label{continuous_defs_section}

For systems with a continuous symmetry, the Hamiltonian  (with the free boundary conditions) is of the form
\begin{equation}
H_{\Gamma,0}^{\vec{h},,\epsilon,\vec{\eta}}=\sum_{A \subset \Gamma} Q_A + \sum_{x \in \Gamma} (\vec{h}+\epsilon \vec{\eta}_x) \cdot \vec{\lop}_x,
\end{equation}
where the $Q$ interaction is not only translation invariant but also rotation invariant in the following sense.  We assume that the group of rotations  $SO(N)$ has a representation by local unitary operators whereby for each $R\in SO(N)$ there is a corresponding $\widehat R_x$ acting in $\mathcal H_x$.
The interaction $Q$ then is rotation invariant if and only if 
\begin{equation}\label{vecHam}
Q_A = \left( \prod_{x \in A} \widehat R^{-1}_x \right) Q_A \left( \prod_{x \in A} \widehat R_x \right)
\end{equation}
for all finite $A \subset \Z^d$ and all $R \in SO(N)$.
At each site $x\in \Z^d$ instead of a single operator $\lop _x$ there is now a collection of operators $\vec{\lop}_x = (\lop_{x,1}, ..., \lop_{x,N})$ which under the above action of $SO(N)$ transform as a vector.  That is: for each $R\in SO(N)$ and $\vec{v}\in \R^N$ 
\begin{equation}\label{vector_operator_transformation}
\widehat R_x^{-1} ( \vec{v} \cdot \vec{\lop}_x  ) \widehat R_x \ = \  (R \vec{v}) \cdot  \vec{\lop}_x \, .  
\end{equation}
Hamiltonians with other boundary conditions, Gibbs states, etc. are defined in analogy with the previous section.   (In the classical case~\cite{AW.CMP} it was also necessary to assume that the interactions transformed smoothly under nonuniform rotations, however for a finite-dimensional local state space this is automatically true.)  
 
Likewise, the random fields are now given by random vectors $\vec{\eta}_x$ in $\R^N$.   Clearly rotation invariance is lost in the presence of such terms in the Hamiltonian.  
However, we will assume that the symmetry is retained in the distributional sense, i.e., that for any rotation $R$ the random variables $\vec{\eta}_x$ and $R \vec{\eta}_x$ have the same distribution.

\section{Statement of the main results} 

\subsection{Two perspectives on \1st order phase transitions}

As was  done in the context of the examples of Section~\ref{WhatIsRounding}, the general statements will  be presented in two equivalent ways:  in their thermodynamic formulations, which is the level at which the results are derived here, and then in the language of statistical mechanics, i.e.\ properties of Gibbs states. 
     Since the results concern (the absence of) first order phase transitions let us recall the latter's dual manifestations.  
 \begin{enumerate} 
 \item In thermodynamic terms, a \1st   order phase transition is associated with the discontinuity of the first derivative of the free energy with respect to one its  parameters.  By default this parameter will be denoted here by $h$. 
 \item  In the terms of statistical mechanics a \1st   order phase transition is expressed   in the non-uniqueness, among the infinite volume Gibbs equilibrium states, of the bulk density of some extensive quantity.  Our discussion concerns the case when this is the quantity 
 whose coupling parameter in $H$ is the field $h$ which is randomized by the disorder. 
 \end{enumerate} 
One  occasionally finds that  the equivalence of these two statements is not fully trusted  in the present context (of disordered systems)\cite{Fisch07}.  We shall therefore briefly recall below, in  Section~\ref{Thermo_and_StatMech}, some pertinent known results.

\subsection{Thermodynamic formulation}\label{sec:main_results}

Our first result concerning systems of the type described in Section~\ref{general_defs}, with $\eta_{x}$  given by independent and identically distributed (i.i.d.) random variables, is:

\begin{theorem} \label{alt_main_prop}
In dimensions $d \le 2$, assuming the variables $\eta_x$ are i.i.d.\ with absolutely continuous  distribution and 
with more than two finite moments, the quenched free energy  density $\F$ is differentiable in $h$ at all values of $h$, $\epsilon\ne 0$, and $\beta \le \infty$.   

Furthermore, for $\beta < \infty$, the 
assumption of absolute continuity can be relaxed, requiring instead that  
  the distribution of $\eta$  has no isolated point masses, or alternatively that 
 the system satisfies the weak FKG property with respect to $\lop$.   
\end{theorem}

Let us recall that a probability  distribution of a variable $q$ is said to be \emph{absolutely continuous} if it is of the form $\rho(q) dq $, with a density function   $\rho(q) $.  In particular, the corresponding measure on $\R$ has neither `delta function' terms (point masses), nor any component  which is supported on a Cantor type fractal set. Point masses are isolated if they are separated from the continuous part of the distribution.
The statement that the random variables $\eta_x$ have more than two finite moments means that
\begin{equation}
\Av | \eta_x |^{2+\delta} < \infty
\end{equation}
for some $\delta > 0$, where $\Av$ denotes the average over the random fields.

As in the classical case, in the presence of continuous symmetry, as understood above,  the `rounding effect'  extends  to higher dimensions:

\begin{theorem} \label{alt_continuous_prop}
In dimensions $d \le 4$, any isotropic system of the type described in Section~\ref{continuous_defs_section}, with the random terms being independent with an identical, rotation invariant distribution with more than two finite moments and 
$\textup{Prob}( \vec{\eta} =\vec{0}) =0$, 
 the quenched free energy density $\F$ is differentiable in $\vec{h}$ at  $\vec{h}=\vec{0}$, for any $\epsilon \ne 0$ and $\beta \le \infty$.

For $\beta < \infty$, the conclusion still holds if there is a nonzero probability of a random term being zero.
\end{theorem}
We note that an apparently weaker condition on the distribution of $\vec{\eta}$ compared to Theorem \ref{alt_main_prop} is adequate because what will ultimately be important is the distribution of the component in an arbitrary direction.  With the assumption of an isotropic distribution for the vector, the components can easily be seen to satisfy the stronger conditions used in Theorem~\ref{alt_main_prop} or Proposition~\ref{AW_theta}.

\subsection{Statistical mechanical implications (no long range order)}\label{Thermo_and_StatMech}  

In this section, we establish the relationship between phase transitions (understood in terms of non-differentiability of free energy) and long range order, culminating in corollaries of Theorems \ref{alt_main_prop} and \ref{alt_continuous_prop} which reframe the results in statistical mechanics terms.

We say that a system exhibits long range order with order parameter $\lop$ (for some set of parameters and disorder variables) if the mean value of the bulk averages  
$ \bar{\lop}_\Lambda = |\Lambda|^{-1} \sum_{x \in \Lambda} \lop_x$, for   $\Lambda \to \Zd$,   depends on the boundary conditions $B$ with which the infinite volume state is constructed.   That is: if there are two sequences of cubic domains $\Lambda$ and $\Gamma$  increasing to $\Zd$, and two sets of boundary conditions which yield different values for 
\begin{equation}
\lim_{\Lambda \nearrow \Zd} \lim_{\Gamma \nearrow \Zd} \state{\bar{\lop}_\Lambda}_{\Gamma,B}.
\label{LRO_limit}
\end{equation}

When the sensitivity to the boundary conditions is found  for the full bulk averages, i.e. if different values occur  for
\begin{equation}
\lim_{\Gamma \nearrow \Zd} \state{\bar{\lop}_\Gamma}_{\Gamma,B},
\label{LLRO_limit}
\end{equation}
we say that there is long - long range order.

\noindent{\bf Remark:} 
If for a given system there are multiple limits of the expectation values in 
\eqref{LRO_limit}, then the system has multiple infinite volume KMS states~\cite{BratelliRobinson2}.   Convex combinations of the different states (which will also form KMS states) will exhibit non-decaying correlations of $\lop$:  $\langle \lop_{x} \lop_{y} \rangle - \langle \lop_{x} \rangle  \langle \lop_{y} \rangle  \not\to 0$ as $|x-y|\to \infty$. 
The latter condition provides another aspect of (``short''-) long range order, which is often used as its definition.

The basis of the connection to the thermodynamic quantities discussed in the previous section is the relation 
\begin{equation}
\frac{\partial \F^h_{\Gamma,\bc}(\eta)}{\partial h} = \frac{1}{|\Gamma|} \sum_{x \in \Gamma} \state{\lop_x}^h_{\Gamma,\bc}(\eta).
\end{equation}
A useful implication of convexity is that if a sequence of differentiable convex functions, such as  $\F$ of  Theorem~\ref{F_converges}, converges pointwise (in $h$) then also their derivatives  converge to the derivative of the limiting function, wherever that function is differentiable.  Without assuming differentiability of the limiting function one may still conclude that all the derivatives' accumulation points lie in the interval spanned by the left and right derivative of the limiting function.   
(An elementary proof of that can be obtained by considering the relations among quotients of the form $[\F(h_2)-\F(h_1)]/[h_2-h_1]$ for suitably chosen collections of intervals $[h_1,h_2]$.)  This has the following relevant implication. 
 
\begin{proposition} \label{LongRange}  Under the assumptions of Proposition~\ref{F_converges}, 
 for any set of the parameters $(\beta,h,\varepsilon)$ at which $\F$ is differentiable in $h$ 
\be  \label{diff}
\lim_{L \to \infty}  \frac{1}{|\Gamma_L|} \sum_{x \in \Gamma_L} \state{\lop_x}^h_{\Gamma_L,\bc}(\eta)  
\ = \  \frac{\partial \F}{\partial h}\,  
\ee 
for almost every realization of the disorder $\eta$, and any choice of the boundary conditions $B$.   Furthermore, also: 
\begin{equation} \label{eq:55}
\lim_{L \to \infty} \lim_{K \to \infty} \frac{1}{|\Gamma_L|} \sum_{x \in \Gamma_L} \state{\lop_x}^h_{\Gamma_K,\bc}(\eta)  \  =\    \frac{\partial \F}{\partial h}\,. 
\end{equation}
Without assuming differentiability, one may still conclude that 
\begin{equation}  \label{diff2}
\LIM_{ L, K \to \infty\,; \, K\ge L } \frac{1}{|\Gamma_L|} \sum_{x \in \Gamma_L} \, \, \state{\lop_x}^h_{\Gamma_L,\bc}(\eta)
\in \left[ \frac{\partial \F}{\partial h-}, \frac{\partial \F}{\partial h+} \right],
\end{equation}     
with $\LIM$ denoting the collection of accumulation points, for different boundary conditions, and possibly different sequences of volumes, and $\tfrac{\partial}{\partial h \pm}$ denoting one-sided derivatives with respect to $h$.  
\end{proposition}
Of particular relevance for us is the conclusion that when $\F$ is differentiable  there is only one possible value for the limit~\eqref{LRO_limit}, which by Proposition~\ref{F_converges} is independent of the boundary conditions.
For classical systems more can be said: if $\F$ is differentiable not only do the mean values of the 
observable $\bar{\lop}_\Gamma$  converge, but the distribution of this quantity with respect to the Gibbs state collapses onto a point.  Such a stronger statement concerning quantum fluctuations is not known to be true.  Nevertheless,   Equation~\eqref{diff} holds also for the quantum systems.

\begin{proof}
The relations~\eqref{diff} and \eqref{diff2} follow directly by the above stated property of convex functions (i.e., the implication for the derivatives of convex functions' pointwise convergence).  

To prove \eqref{eq:55} and the corresponding extension of \eqref{diff2}, 
let $F^{h,\delta,\Lambda}_{\Gamma,B}$ denote the free energy with the fixed field within $\Lambda$ changed by $\delta$, so that
\begin{equation}\label{SRO_prop_derivs}
\frac{1}{|\Gamma_L|} \sum_{x \in \Gamma_L} \state{\lop_x}^h_{\Gamma_K,\bc}(\eta) = \frac{1}{|\Gamma_L|} \left.\frac{\partial F^{h,\delta,\Lambda}_{\Gamma_K,B}}{\partial \delta} \right|_{\delta=0}
\end{equation}
A standard estimate, which is presented in Proposition~\ref{Ruelle_lemma} below,  implies that
\begin{equation}
F^{h,\delta,\Gamma_L}_{\Gamma_K,B} - F^{h}_{\Gamma_K,B} 
= F^{h+\delta}_{\Gamma_L,0} - F^{h}_{\Gamma_L,0} + O \left(  \|V_{\Gamma_L}^{\eta}\| \right)
\end{equation}
uniformly in $K$, where $V^{\eta}_L $ denotes the terms in the Hamiltonian crossing the boundary of $\Gamma_L$, i.e.
\begin{equation} \label{Vdef}
V_{\Gamma}^\eta := \sum_{A: \overlappingClass{A}{\Gamma}} Q_A + \sum_{x\in \partial_0 \Gamma} (h + \epsilon \eta_x) \lop_x .
\end{equation} 
Then for any $\eta$ in the full measure set $\mathcal{N}$ on which the limits in Theorem~\ref{F_converges} are defined, this implies that
\begin{equation}
\lim_{ L, K \to \infty\,; \, K\ge L }   \left( F^{h,\delta,\Gamma_L}_{\Gamma_K,B} -  F^{h}_{\Gamma_K,B} \right) \, /\, |\Gamma_L| 
=\F(h+\delta,\beta)-\F(h,\beta).
\end{equation}
each of these differences is a concave function of $\delta$, so the  above mentioned convexity argument gives
\begin{equation}
\LIM_{ L, K \to \infty\,; \, K\ge L }  \, \frac{\partial}{\partial \delta}  F^{h,\delta,\Gamma_L}_{\Gamma_K,B} \, /\, |\Gamma_L| \ 
\in \   \left[ \frac{\partial \F}{\partial h-}, \frac{\partial \F}{\partial h+} \right]
\end{equation}  
and the desired results follow immediately from Equation~\eqref{SRO_prop_derivs}.  
\end{proof}

%
%
 In particular,  the results which are formulated in Theorems~\ref{alt_main_prop} and \ref{alt_continuous_prop}  as statements of differentiability of the free energy directly imply the following.

\begin{theorem} \label{main_prop} \label{continuous_prop}
In dimensions $d \le 2$, under the assumptions of Theorem~\ref{alt_main_prop} the system almost certainly does not  exhibit (short or long) long range order.   

Furthermore, for systems with continuous symmetry which meet the assumptions of Theorem~\ref{alt_continuous_prop}, in dimensions $d \le 4$  almost certainly there is no (short or long) long range order at~$\vec{h} = \vec{0}$. 
\end{theorem}

\section{The free-energy-difference functional}

The above discussion reduces our main results to Theorems~\ref{alt_main_prop} and \ref{alt_continuous_prop}.   
A key tool for the proof of these theorems  is a quantity which describes the differences in the finite volume content of free energies between two of the system's equilibrium states, which are constructed to have the two extremal mean magnetizations, $ \frac{\partial \F}{ \partial h+}$ and $ \frac{\partial \F}{ \partial h-}$, assuming they are not the same.  The difference, which is denoted below as $G_L$, will be shown to satisfy contradictory bounds:  
{\it i.} an absolute upper bound on $|G_L|$ which is due to the observation that the two free energies cannot differ by more than the magnitude of the interaction across the finite volume boundary, {\it ii.}  a lower bound on the fluctuations of this quantity   which reflects the idea that with a systematic difference in the magnetization the states'  free energies will respond differently to the fluctuating random fields.  

More specifically,  the free energy differences (properly defined) will be shown to satisfy: 
\begin{equation}\label{simpleUBound}
  |G_L| \le C L^{d-1} + D L^{d/2}  \, ,  
\end{equation}
with $D=0$ for finite range interactions.  In the presence of continuous symmetry for  finite range interactions the bound is improved to:
\begin{equation}\label{continuousUBound}
  |G_L| \le C L^{d-2} \, .
\end{equation}
For  the opposite bound it will be shown that  if there is a first order transition then $G_L / L^{d/2}$ converges to a normally distributed random variable with variance $b > 0$: 
\begin{equation}\label{LBound}
  \frac{G_L}{L^{d/2}} \rightarrow  N(0,b) \,. 
\end{equation}
The construction of a quantity with the above characteristics  is the main subject of this section.     The contradiction between the two bounds  yields the main result.

Estimates similar to the above form the basis of the argument of Imry and Ma\cite{YM}, and the more precise  argument outlined above is similar to the one which was employed for the analysis of in the classical case~\cite{AW.PRL, AW.CMP}.  However the technique introduced there for the construction of the auxiliary quantity $G_L$ employed probability measures over states, defining what has since been called metastates~\cite{Newman_Stein}.  Unfortunately, although metastates can be defined for quantum systems~\cite{Barreto, *Fidaleo}, not all the steps taken in  Ref.~\onlinecite{AW.CMP} for the construction of $G_L$   have such an extension.
The progress made in this work is enabled by the observation that one can define a suitable quantity $G_L$ in terms of just free energy differences, avoiding more delicate issues of quantum states.   This makes it possible to formulate a proof parallel to the classical case (and indeed the nonrigorous arguments which inspired it).

As a simple illustration of the concept one may first consider the zero temperature case of a system such as the random field Ising model with two  distinct states, labeled by $+$ and~$-$, of different mean magnetizations and with only the `canonical dependence' on  $\eta$ (cf. \cite{AW.CMP}).  In that case,  for  $G_L$ one may take   the difference in the two states' energy contents in finite volumes: 
\begin{equation}
\begin{split}
E_L(\eta)\ &:= \ E_{+,L}(\eta) - E_{-,L}(\eta)  = 
 \epsilon \state{\sum_{x \in \Gamma_L} \eta_x \lop_x}_+ - \epsilon \state{\sum_{x \in \Gamma_L} \eta_x \lop_x}_- \\ & \approx  2 \epsilon \left( \sum_{x \in \Gamma_L} \eta_x \right) M \ \to 2 \ \epsilon M L^{d/2} N(0,1),
\end{split}
\end{equation}
where the last step, in which the central limit theorem is invoked, is valid only to the extent that one may ignore the adaptation of the system to the disorder.  (A more precise statement is possible taking into account the dependence of the $\pm$ states on $\eta$~\cite{AW.CMP}).   
The above  provides only a suggestive example in the context of a special (and classical) system at $T=0$.  We  turn now to a more general definition which would be suitable for our purpose.



Our choice of $G_L(\eta)$ for the general case is based on the following  free energy `second difference' (in the field $h$, and disorder $\eta_{\Gamma_L}$):  
\begin{equation}\label{GdeltaDef}
\gf{L,K}{\delta}(\eta_L) := \frac{1}{2} \condAv{F^{h+\delta}_K (\eta) - F^{h+\delta}_K(r_L(\eta)) - F^{h-\delta}_K (\eta) + F^{h-\delta}_K(r_L(\eta))}{\eta_{L}},
\end{equation}
where  $r_L{\eta}$ is the field obtained by setting $\eta_x=0$ for all $x \in \Gamma_L$, $\eta_L$ is the restriction of $\eta$ to $\Gamma_L$ and $\condAv{\cdot}{\eta_L}$ is the conditional average over  the random terms outside $\Gamma_L$, i.e.  with $\eta_{\Gamma_L}$ held constant.   It should be noted that $F_K$ (which is defined in \eqref{eq:per}) refers  to the free energy with the periodic boundary conditions.  This choice  assures translation covariance, which is mentioned below (condition 4. in Lemma  \eqref{th:convergence}) and used in the argument. 
  
When the double limit $\lim_{\delta\searrow 0} \lim_{K\to \infty} \gf{L,K}{\delta}(\eta_L) $, exists (for all $\eta_L$), the quantity defined by it has  properties we desire of $G_L(\eta_L)$.   (The order of the limits is important here: taking $\delta \searrow 0$ after the infinite volume limit serves to construct the general version of the  $+$ and $-$ states.)  
Inconveniently, the limits are not generally known (or expected in full generality) to  exist.  However compactness arguments can be applied to prove sufficient convergence along subsequences.   The essential statement is the following.   We use here the  $\ell^1$-Lipschitz seminorm $\lip{\cdot}$  
 (in lieu of a uniform bound on the derivative) which is defined as 
\begin{equation} \label{lip1def}
\lip{f} :=
\sup_{\substack{\eta,\eta' \in \Fields \\ 0<\pnorm{\eta-\eta'}{1} < \infty}} \frac{|f(\eta)-f(\eta')|}{\pnorm{\eta-\eta'}{1}} \,  ,
\end{equation}
with 
$\pnorm{\eta}{1} := \sum_{x \in \Zd} |\eta_x|  \, . $
  
%
%

\begin{lem}\label{th:convergence}
  For any $\beta \le \infty$, there are sequences $K_j \to \infty$ and $\delta_i \to 0$ such that the limits
  \begin{equation}
G_L(\eta_L) := \lim_{i \to \infty} \lim_{j \to \infty} \left( \gf{L,K_j}{\delta_i} - \Av \left[ \gf{L,K_j}{\delta_i} \right] \right)
\end{equation}
exist for all $L$ and $\eta$, and have the following properties:
\begin{enumerate}
\item $\Av G_L(\eta_L) = 0$ \label{ct_zero_mean}
\item $G_L(\eta_L)$ depends on the values of $\eta_x$ only for $x \in \Gamma_L$ \label{ct_local}
\item $\lip{G_L} \le \epsilon$ \label{ct_lip}
\item $\condAv{G_L}{\eta_\Lambda} = G_{L'}( T_x \eta)$ whenever $T_{-x} \Gamma_{L'} = \Lambda \subset \Gamma_L$ \label{ct_covariance}
\item $G_1$ (that is, $G_L$ with $L=1$, which is a function of one variable) has a distributional derivative $G_1'$ satisfying $\lip{\frac{\partial G_1}{\partial \eta_0}} \le \beta \epsilon^2$ \label{ct_lip2} and
\item  $ \Av G_1'(\eta_0) = \frac{\epsilon}{2} \left( \frac{\partial \F}{\partial h+} - \frac{\partial F}{\partial h-} \right)$ \label{ct_M}
\end{enumerate}
Furthermore, if the system satisfies the weak FKG condition with respect to $\lop$, then $G_1$ is monotone.  For $\beta=\infty$ the same results hold assuming that the distribution of $\eta$ is absolutely continuous.
\end{lem}

The proof is postponed to section~\ref{main_proofs}, as it is based on compactness arguments paralleling those in Ref.~\onlinecite{AW.CMP}.  Before turning to it, in the next two sections we will show that any such $G_L(\eta_L)$ must obey the upper bound~\eqref{simpleUBound}, and where appropriate~\eqref{continuousUBound}, and also satisfies~\eqref{LBound}.  


\section{Upper bounds on the free energy difference}

We will frequently use the following estimate to control the effects of changes in the Hamiltonian on the free energy. 
\begin{proposition}[{\cite{Ruelle}}]\label{Ruelle_lemma}
For any Hermitian matrices $C,D$ of the same finite size,
\begin{equation} \label{eq:FreeEnergyDifferences}
\left| \log \Tr e^C - \log \Tr e^D \right| \le \|C-D\|
\end{equation}
where $\| \cdot\|$ is the operator norm.
\end{proposition}
As a simple application, this means that the change in free energy for the system defined by the Hamiltonian~\eqref{eq:H} when boundary conditions are added is bounded in terms of the total of the norms of all terms in the interaction which cross the boundary.   
 
\subsection{A general surface bound}\label{sec:surface_FE_bound}

The following lemma shows generally that the free energy difference due to random field fluctuations in a finite region can be bounded by the norm of the interaction terms crossing the surface of that region, and that for short range interactions this is proportional to the area of that surface.
\begin{lem}\label{simple_UBound_lemma} 
Assuming the interactions  are either of finite range or satisfy  \eqref{eq:decay} (of Appendix \ref{LR_appendix}), 
for any pair of sequences $\delta_i \to 0$, $K_j \to \infty$ for which the following sequence of functions converges for all $\eta_{L}$
  \begin{equation}
G_L(\eta_L) = \lim_{i \to \infty} \lim_{j \to \infty} \left( \gf{L,K_j}{\delta_i} - \Av \left[ \gf{L,K_j}{\delta_i} \right] \right)
\end{equation}
the limiting function satisfies  (with some  $C,D < \infty$):
\begin{equation}\label{basic_upper_bound1}
  |G_L| \le C L^{d-1} + D L^{d/2}  \, . 
\end{equation}
\end{lem}
\begin{proof}
We will show that
\begin{equation}
   |\gf{L,K}{\delta}| \le c L^{d-1} + b L^{d/2} + O(\delta L^d)
\end{equation}
with $c$, $b$ independent of $L$, $K$, and $\delta$.  The desired result holds then with $C=2c$, $D=2b$.

To see this, we let $\Lambda_L$ be the smallest cube in $\Zd$ so that $\lop_x$ acts within $\Lambda_L$ for all $x \in \Gamma_L$, and then split the Hamiltonian into contributions from $\Lambda_L$ and $\Gamma_K \setminus \Lambda_L$ and terms connecting the two.  We consider the quantity
\begin{equation}\label{split_F}
F^h_{K | L} (\eta) := - \frac{1}{\beta} \log \Tr \exp \left( - \beta \left[ H_{\Lambda_L,0}^{h,\eta} + H_{\Gamma_K \setminus \Lambda_{L, *}}^{h,\eta}\right]\right),
\end{equation}
where the subscript $0$ refers to free boundary conditions, and the subscript $*$ refers to periodic boundary conditions on the edge of $\Gamma_K$ and free boundary conditions on the edge of $\Lambda_L$.
The Hamiltonian in Equation~\eqref{split_F} differs from $H^{h,\eta}_{\Gamma_K}$ only in the absence of terms connecting $\Lambda_L$ to the rest of $\Gamma_K$, which as in  Equation~\eqref{Vdef} we denote by $  V_{\Lambda_L}^{\eta}$.  Then by Proposition~\ref{Ruelle_lemma}
\begin{equation}
\left| F^h_{K | L} (\eta) - F^h_{K} (\eta) \right| \le \left\| V_{\Lambda_L}^{\eta} \right\|,
\end{equation}
and therefore
\begin{equation}
\begin{split}
& \left| F^{h+\delta}_K (\eta) - F^{h+\delta}_K(r_L(\eta)) - F^{h-\delta}_K (\eta) + F^{h-\delta}_K(r_L(\eta)) \right|
\\ & \ \
\le \left| F^{h+\delta}_{K | L} (\eta) - F^{h+\delta}_{K | L}(r_L(\eta)) - F^{h-\delta}_{K | L} (\eta) + F^{h-\delta}_{K | L}(r_L(\eta)) \right|
+ 4 \left\| V_{\Lambda_L}^{\eta} \right\|.
\end{split} \label{Boundary_estimate_1}
\end{equation}

Each $F_{K | L}$ is a sum of a part from $\Lambda_L$ and a part from the rest of $\Gamma_L$; the latter cancel each other exactly, and the former can be arranged in two pairs which differ only in the constant field term which has norm of order $\delta L^d$.
Taking the average conditioned on $\eta_{\Gamma_L}$ and noting that the definition of $\Lambda_L$ makes $V_{\Lambda_L}^{\eta}$ independent of $\eta_{\Gamma_L}$,
\begin{equation}
\left| \gf{L,K}{\delta}(\eta_L) \right| \le 2\Av \left\| V_{\Lambda_L}^{\eta} \right\| + O(\delta L^d)
\end{equation}
for all $\eta$,
 and since the $\delta$ term is uniform in $K$,
\begin{equation}\label{G_L_upper_bound}
|G_L(\eta_L)| \le 2 \Av \left\| V_{\Lambda_L}^{\eta} \right\|.
\end{equation}
This reduces Inequality \eqref{basic_upper_bound1} to a comparable bound on $ V_{\Lambda_L}^{\eta}$, i.e.\ 
\begin{equation} \label{needed_boundary_estimate}
\Av \left\| V_{\Lambda_L}^{\eta} \right\| \le c L^{d-1} + b L^{d/2}.
\end{equation}

For finite range interactions it is obvious that 
$\Av \left\| V_{\Lambda_L}^{\eta} \right\| \le c L^{d-1} $, 
and Appendix~\ref{LR_appendix} discusses more general conditions under which Inequality \eqref{needed_boundary_estimate} holds.
\end{proof}

\subsection{An improved bound for systems with continuous symmetry}

For systems with continuous symmetry the stronger bound of Inequality~\eqref{continuousUBound} can be obtained through the analysis of soft deformations.  The argument is similar to the one which was carried out in the classical context in Ref.~\onlinecite{AW.CMP}.   Although the estimate may at first sight appear  to be simply a calculation of the Bloch spin wave energy, that  alone would not  yield the desired result since the first order term is potentially much larger than the required bound.   However, by comparing the distortion energies of  opposite deformations one finds that in any situation for one of the distortions the first order term is of the desired sign.  Therefore the analysis can be limited to the second order term - and that  yields the claimed bound.

We need to modify the definition of $G_L$ slightly to accommodate random vector fields.  The argument will be based on the effect of the fluctuations of a single component, described by expressions where the others have been averaged out.  We let
\begin{gather}
\gf{L,K}{\delta\hat{e}}(\eta_L) = \frac{1}{2} \condAv{
F^{\delta\hat{e}}_K(\vec{\eta}) - F^{\delta\hat{e}}_K(r_L(\vec{\eta}))
- F^{-\delta\hat{e}}_K(\vec{\eta})
+ F^{-\delta\hat{e}}_K(r_L(\vec{\eta}))
}{\hat{e} \cdot \vec{\eta}_L}
\\[2ex] 
G_L(\eta_L) = \lim_{i \to \infty} \lim_{j \to \infty} \left( \gf{L,K_j}{\delta_i \hat{e}}(\eta)- \Av \left[  \gf{L,K_j}{\delta_i \hat{e}}(\hat{e} \cdot \vec{\eta}) \right] \right)
\end{gather}
where $\hat{e}$ is an arbitrary unit vector; note that the rotation symmetry of the system and of the distribution of $\vec{\eta}$ mean that the right hand sight of the last expression is independent of $\hat{e}$.
 
\begin{lem}
  For any system with continuous symmetry and isotropic disorder as described in Section~\ref{defs_section}, of interactions which are either of finite range or satisfy  \eqref{continuous_short_range_assum}, at  $\vec{h}=0$:
 \begin{equation}  \label{eq:77}
  |G_L(\eta_L)| \le C L^{d-2}  \, , 
\end{equation}
with $C<\infty$,  uniformly in $L$ and $\eta_{L}$.
\label{th:continuous_upper_bound}
\end{lem}
\begin{proof}
We shall derive \eqref{eq:77} through uniform bounds  on
\begin{equation} \label{g_Kini_def}
g_{L,M}^{\delta \hat{e}} (\vec\eta) := \condAv{F_K^{\delta \hat{e}}(\vec{\eta}) - F_K^{-\delta \hat{e}}(\vec\eta) }{\vecEtaCond{L}} \, ,  
\end{equation} for which we shall show that Assumption~\eqref{continuous_short_range_assum} implies
\begin{equation} \label{eq:99}
|g_{L,K}^{\delta \hat{e}} (\vec\eta)| = O(L^{d-2})  \, .  
\end{equation}
The two quantities are related through
\begin{equation}\label{G_and_g}
G_L(\eta_L) = \frac{1}{2} \lim_{i \to \infty} \lim_{j \to \infty} \left( g_{L,K_j}^{\delta_i \hat{e}} (\eta) - \Av g_{L,K_j}^{\delta_i \hat{e}} (\hat{e} \cdot \vec\eta) \right),
\end{equation}
and hence \eqref{eq:99} implies the desired statement.  


Let $\rho$ be the generator (in $so(N)$) of a rotation in a plane containing $\hat{e}$, and for each $x \in \Zd$ let $\rho_x$ be the generator of the corresponding rotation in the single-site algebra $\alg_x$ (Unlike in Ref.~\onlinecite{QIMLetter}, we will use the ``mathematician's'' convention that rotations are given by $e^{\theta \rho}$, so that $\rho$ is an antihermitian matrix and $\rho_x$ is an antihermitian operator).  We introduce the slowly varying angles
\begin{equation}\label{angles_theta_def}
\theta_x := \piecewise{
0, & x \in \Gamma_L \\
\frac{\|x\|_1-L}{L}\pi, & 0 < d_L(x) < L  \\
\pi, &  d_L(x) \ge L},
\end{equation}
where $d_L(x)$ is the distance from $x$ to $\Gamma_L$ in the largest-component metric, i.e.\ \begin{equation}
 d_L(x):= \min_{y \in \Gamma_L} \pnorm{x-y}{\infty}.
\end{equation}
We also introduce the associated rotations on fields and on $\alg_\Gamma$ defined by
\begin{gather}
\widehat{R}_x := e^{\theta_x \rho} \\
\left(R_\theta(\vec \eta)\right)_x \equiv R_x \vec \eta_x \\
\widehat{R}_\theta = \bigotimes_{x \in \Gamma} e^{\theta_x \rho_x}.
\end{gather}
where $\Gamma$ is the relevant finite subset of $\Zd$ which can be inferred from the context.

$\widehat{R}_\theta$ is unitary, and so we can rewrite the free energy $F_K^{-\delta \widehat{e}}(\vec\eta) $ appearing in~\eqref{g_Kini_def} as
\begin{equation} \label{rotateH}
F_K^{-\delta \widehat{e}}(\vec\eta) = -\frac{1}{\beta} \log \Tr \exp\left( -\beta \widehat{R}_\theta^{-1} H_{\Gamma_K}^{-\delta \widehat{e},\vec{\eta}} \widehat{R}_\theta  \right).  
\end{equation}
At this stage we could carry out a spin-wave analysis, writing
\begin{equation}
\begin{split}
&\condAv{F_K^{-\delta \widehat{e}}(\vec\eta)}{\vecEtaCond{L}}
= -\frac{1}{\beta} \condAv{\log \Tr \exp\left( -\beta [ H_{\Gamma_K}^{\delta \widehat{e},\vec{\eta}} + \Delta H_\theta ] \right)}{\vecEtaCond{L}},
\end{split}
\end{equation}
which by Lemma~\ref{Ruelle_lemma} implies
\begin{equation}
|g_{L,K}^{\delta \widehat{e}} (\vec\eta)| \le \|\Delta H_\theta\|  \, . 
\end{equation}
However this will not lead to the desired result, since 
$\|\Delta H_\theta\|$ includes  `first order' terms (in $\nabla \theta $) which  scale as $L^{d-1}$ as well as  the  `second order' terms (and higher) which scale as $L^{d-2}$.  To cancel the former  we rewrite $F_K^{-\delta \widehat{e}}(\vec\eta)$ using an opposite rotation and average the two expressions, to obtain
\begin{widetext}
\begin{equation}\label{F_bounds_for_g}
\begin{split}
\condAv{F_K^{-\delta \widehat{e}}(\vec\eta)}{\vecEtaCond{L}} =  -\frac{1}{2 \beta} \Av & \left[ \log \Tr \exp\left( -\beta [ H_{\Gamma_K}^{\delta \widehat{e},\vec{\eta}} + \Delta H_\theta ] \right) \right.
\\ & \left. + \log \Tr \exp\left( -\beta [ H_{\Gamma_K}^{\delta \widehat{e},\vec{\eta}} + \Delta H_{-\theta} ] \right) \middle| \vecEtaCond{L}\right].
\end{split}
\end{equation}

A combination of the Cauchy-Schwarz inequality, the Golden-Thompson inequality~\cite{Bhatia} and Lemma~\ref{Ruelle_lemma} yields the general relation  
\begin{equation}
\begin{split}
\log \Tr e^A  - \frac{1}{2}\log \Tr e^B - \frac{1}{2}\log \Tr e^{C}
&= \log \left( \frac{ \Tr e^A}{(\Tr e^{B})^{1/2} (\Tr e^{C})^{1/2}} \right)
 \le \log \left( \frac{ \Tr e^A}{\Tr e^{B/2} e^{C/2}} \right)
\\[2ex]  &\le \log \Tr e^{A-(B+C)/2} \le \left\| A- \frac{B+C}{2} \right\|
\end{split}
\end{equation}
\end{widetext}
for arbitrary Hermitian matrices $A,B,C$.  Applying this to Equation~\eqref{F_bounds_for_g} gives
\begin{equation}\label{g_upper_bound}
g_{L,K}^{\delta \widehat{e}} (\vec\eta) \le \frac{1}{2} \left\| \Delta H_\theta + \Delta H_{-\theta} \right\|.
\end{equation}

We now write out the terms in the rotated Hamiltonian in Equation~\eqref{rotateH} as
\begin{equation}
\begin{split}
&\widehat{R}_\theta^{-1} H_{\Gamma}^{h,\vec{\eta}} \widehat{R}_\theta \\ &= \widehat{R}_\theta^{-1} \left( \sum_A P_\Gamma(Q_A) + \sum_{x \in \Gamma} (-\delta \widehat{e}+\epsilon \vec\eta_x) \cdot P_\Gamma(\vec{\lop}_x)
\right) \widehat{R}_\theta,
\end{split}
\end{equation}
where $ P_\Gamma(Q_A)$ and $ P_\Gamma(\vec{\lop}_x$ denote the terms appearing in the Hamiltonian with periodic boundary conditions, obtained by mapping each site outside $\Gamma$ to the corresponding site within $\Gamma$.

Since $\vec{\lop}$
are vector operators (recall Equation~\eqref{vector_operator_transformation}),
\begin{equation}
\vec \eta_x \cdot \left( \widehat{R}_\theta^{-1} \vec \lop_x \widehat{R}_\theta \right) = \vec \eta_x \cdot R_\theta(\vec \lop)_x = \left[R^{-1}_\theta(\vec \eta)_x\right] \cdot \vec \lop_x;
\end{equation}
inside $\Gamma_L$ there is no rotation, and outside we are performing an average with respect to an isotropic distribution, so this term makes no contribution to $\Delta H$.

As for the fixed field terms, we have
\begin{equation}
\widehat{e} \cdot \left( \widehat{R}_\theta^{-1} \vec \lop_x \widehat{R}_\theta \right) = \left( R_x^{-1} \widehat e \right) \cdot \vec\lop_x.
\end{equation}
The choices of $\rho$ and $\theta$ were intended precisely to make $R_x \widehat e = -\widehat e$ for $d_L(x) > L$; and for the remaining $(3L)^d$ sites we have $\left\| \widehat{e} \cdot \left( \widehat{R}_\theta^{-1} \vec \lop_x \widehat{R}_\theta \right) + \widehat{e} \cdot \vec \lop_x \right\| \le 2$, so these terms make a contribution to $\Delta H$ which is uniformly bounded in norm by $2(3L)^d \delta$.

\begin{widetext}
We are left with the terms arising from the transformation of the nonrandom interaction.  For legibility we let $
\Gamma = \Gamma_K$ for the rest of this section. 
For any $A$ and any (arbitrarily chosen) $x \in A \cap \Gamma$,
\begin{equation}
\begin{split}
\widehat R_{-\theta} P_\Gamma(Q_A) \widehat R_\theta
   &= \left( \bigotimes_{y \in A \cap \Gamma} e^{-(\theta_y - \theta_x)\rho_y} e^{-\theta_x \rho_y} \right) P_\Gamma(Q_A) \left( \bigotimes_{z \in A \cap \Gamma} e^{-\theta_x \rho_z} e^{(\theta_z - \theta_x)\rho_z} \right)
\\ & = \left( \bigotimes_{y \in A \cap \Gamma} e^{-(\theta_y - \theta_x)\rho_y} \right) P_\Gamma(Q_A) \left( \bigotimes_{z \in A \cap \Gamma} e^{(\theta_z - \theta_x)\rho_z} \right),
\end{split}
\end{equation}
(using the rotation invariance of $Q_A$).  Expanding the exponentials, we obtain
\begin{equation}
\begin{split}
\widehat R_{-\theta} P_\Gamma(Q_A) \widehat R_\theta = P_\Gamma(Q_A) &+ \sum_{y \in A \cap \Gamma} (\theta_x - \theta_y)\left(\rho_y  P_\Gamma(Q_A) - P_\Gamma(Q_A) \rho_y \right) \\ & + O\left( \frac{(\diam A)^2|A|^2}{L^2}\left\| Q_A \right\| \right),
\end{split}
\end{equation}
where the estimate of the higher order terms uses
\begin{equation}
|\theta_x - \theta_y| \le \frac{\pi \|x-y\|_\infty}{L} \le \frac{\pi \diam A}{L}
\end{equation}
and the observation that the $n$th order term in the expansion is potentially a sum of $|A|^n$ terms, as well as $\left\| P_\Gamma(Q_A) \right\| \le \left\| Q_A \right\| $.  The first order terms are odd in $\theta$, and will cancel in $\Delta H_\theta + \Delta H_{-\theta}$, with the leading term being second order.  What appears there is
\begin{equation}
\begin{split}
\sum_{A \cap \Gamma \ne \emptyset} \left(\widehat R_{-\theta} P_\Gamma(Q_A) \widehat R_\theta -  P_\Gamma(Q_A) \right) = O\left( L^d \sum_{A \ni 0} \frac{1}{|A|} \frac{(\diam A)^2|A|^2}{L^2} \left\| Q_A \right\| \right)  = O(L^{d-2}), \label{continuous_interaction_bound}
\end{split}
\end{equation}
where for non-finite-range interactions the last equality requires the assumption \eqref{continuous_short_range_assum}, which is presented in Appendix \ref{LR_appendix}.
\end{widetext}

Then the right hand side of Inequality~\eqref{g_upper_bound} is
\begin{equation}
\|\Delta H_\theta + \Delta H_{-\theta} \| = O(L^{d-2}) + O(\delta L^d).
\end{equation}

This provides only an upper bound on $g_{L,M}^{\delta \widehat{e}} (\vec\eta)$, rather than a bound on its absolute value.  However it is obvious from the definition~\eqref{g_Kini_def} of $g$ that $g_{L,M}^{\delta \widehat{e}} (\vec\eta) = - g_{L,M}^{-\delta \widehat{e}} (\vec\eta)$, so the needed lower bound follows automatically.  
This proves~\eqref{g_Kini_def}, and thus, thorough~\eqref{G_and_g}, the claimed bound~\eqref{eq:77}.
\end{proof} 
   
\section{Stochastic lower bounds on the local free energy difference}\label{CLT_section}

To prove the  Imry and Ma\cite{YM} phenomenon, Aizenman and Wehr\cite{AW.CMP} employed a somewhat generalized form of the central limit theorem  which is suitable for the families of random variables presently under consideration~\cite{HallHeyde,AW.CMP}.   We will use the following reformulation of that statement.  Since the proof is in the literature it will be omitted here; one incorporating a correction due to Bovier\cite{Bovier} can be found in Ref.~\onlinecite{RLG.thesis}. 

Recalling that $T_x$ is the operation of translation by $x \in \Zd$ and $\condAv{\cdot}{\eta_\Lambda}$ is the conditional average conditioned on the values of $ \{ \eta_x\}_{x\in \Lambda} =: \eta_\Lambda $,  we present:

\begin{proposition}\label{CLT_prop}
Let $\eta_x$ be a collection of i.i.d.\ random variables (indexed by $x \in \Zd$) with $\Av |\eta_x|^{2+\delta}<\infty$ for some $\delta > 0$, and let $G_L$ be a family of real functions indexed by $L \in \mathbb{N}$, each with the following properties:
\begin{enumerate}
\item $\Av G_L(\eta) = 0$
\item $\lip{G_L} \le \epsilon$ \label{uLif} \quad  (with  $\lip{\cdot}$  the  Lipschitz seminorm  defined by  
\eqref{lip1def})
\item $G_L(\eta)$ depends on the values of $\eta_x$ only for $x \in \Gamma_L$ \label{local_hypothesis}
\item $\condAv{G_L(\eta)}{\eta_\Lambda} = G_{L'}(T_x\eta_\Lambda)$ whenever $T_{-x} \Gamma_{L'} = \Lambda \subset \Gamma_L$ \label{consistency_assumption}
\end{enumerate}
Then 
\begin{equation}
G_L(\eta)/L^{d/2} \,  {\longrightarrow} \, N(0,b)
\end{equation}
in distribution as $L \to \infty$, for some $b$ satisfying
\begin{equation} \label{bBounds}
\Av G_1^2 \le b^2 \le 2 \epsilon^2 (\Av |\eta_0|)^2 \,  .
\end{equation}
\end{proposition}
Assumption~\ref{consistency_assumption} in essence states  that the different functions involved are essentially the same quantity at different scales, and that it is in some sense translation invariant.  
More precisely:   the family forms a  stationary Martingale, in a multidimensional sense of this condition.   

In order to use this result, we will need to establish some control over the conditions under which $\Av{G_1^2} > 0$.  To do this, we employ the following criterion (proven in Appendix III of Ref.~\onlinecite{AW.CMP}):

\begin{proposition} \label{AW_theta}
Let $\nu$ be a Borel probability measure on $\R$, and let $g$ be a continuous function with $\lip{g} = 1$ and 
\begin{equation}
  \int g' (t) \nu(dt) > 0.
\end{equation}

Then any of the following is a sufficient condition for $\int g(t)^2 \nu(dt) > 0$:
\begin{enumerate}
\item $\nu$ is absolutely continuous
\item $g$ is differentiable with $\lip{g'}$ finite, and $\nu$ has no isolated point masses.
\item $g$ is differentiable with $\lip{g'}$ finite, $g' \ge 0$, and $\nu$ is not concentrated at a single point.
\end{enumerate}
\end{proposition}

To apply this we note that $G_1$ is a function of one variable ($\eta_0$) which satisfies the conditions given above.  In particular, it will be shown that the criteria 2.\ and sometimes 3.\ are satisfied at finite temperatures,   in which case $\lip{\frac{\partial G_1}{\partial\eta_0}}$ is uniformly bounded.  In addition, we will see that the weak FKG property implies monotonicity of $G_1$.  

\section{Conclusion of the proofs of the main results}\label{main_proofs}

\subsection{Existence of an infinite-system limit for free energy fluctuations}

Having established the properties we require of the infinite-system free energy fluctuation $G_L$, we now proceed to show that such a suitable quantity exists.  This will be done in two overlapping cases.  We will have occasion to use the following equicontinuity arguments:

\begin{lem} \label{compactness_lemma_1}
Let $f_{ij}:\R^N \to \R$ be a family of functions labeled by $i,j \in \N$, each satisfying $\lip{f_{ij}} \le 1$ and $f_{ij}(0) = 0$.  Then there are subsequences $i_k,j_l$ such that
\begin{equation}
f(z) = \lim_{k \to \infty} \lim_{l \to \infty} f_{i_k j_l}(z)
\end{equation}
exists for all $z \in \R^N$.  Furthermore the convergence is uniform on any compact domain $\Xi \subset \R^N$.
\end{lem}

\begin{lem}\label{compactness_lemma_2}
Let $f_{ij}:\R^N \to \R$ be as in Lemma~\ref{compactness_lemma_1} above, without assuming $f_{ij}(0)=0$.  Then the same result holds assuming instead that there is some $c$ such that $|f_{ij}(z)| \le c < \infty$ for all $z \in \R^N$, $i,j \in \mathbb{N}$.
\end{lem}

These two lemmata are proven by applying the Arzel\`a-Ascoli theorem to obtain compactness of the space of sequences involved, then repeatedly applying the diagonal subsequence trick to obtain a single pair of sequences simultaneously satisfying all of the uniformity criteria.\cite{RLG.thesis}

\subsubsection{Finite temperature}

We now begin with the proof of Theorem~\ref{th:convergence} in the case $\beta < \infty$.  Firstly, note that it is obvious that if the specified limits exist, then they will have the properties labeled~\ref{ct_zero_mean} and~\ref{ct_local} in that theorem.  To define the sequence used we proceed as follows.
For $x \in \Gamma_L$, let
\begin{equation}
\phi^x_{L,K,\delta}(\eta) := \frac{\partial \gf{L,K}{\delta}}{\partial \eta_x} = \frac{\epsilon}{2} \condAv{ \state{\lop_x}^{h+\delta}_K(\eta) - \state{\lop_x}^{h-\delta}_K(\eta)}{\eta_L}; \label{phiDef}
\end{equation}
then evidently
\begin{equation} \label{thetaBound}
| \phi^x_{L,K,\delta}(\eta)| \le \epsilon
\end{equation}
\begin{widetext} and
\begin{equation}
\begin{split}\label{2LBound}
\left| \frac{\partial \phi^x_{L,K,\delta}}{\partial \eta_y} \right| =& \frac{\beta \epsilon^2}{2}  \left| \state{\lop_x \lop_y}^{h+\delta}_K(\eta) - \state{\lop_x \lop_y}^{h-\delta}_K(\eta) \right.
\\ &\left.
- \state{\lop_x}^{h+\delta}_K(\eta)  \state{\lop_y}^{h+\delta}_K(\eta) + \state{\lop_x}^{h-\delta}_K(\eta)  \state{\lop_y}^{h-\delta}_K(\eta)
\right|\leq 2\beta \epsilon^2.
\end{split}
\end{equation}
\end{widetext}

For $\beta < \infty$, this means that for each $L$, $\phi^x_{L,K,\delta}$ is an equicontinuous family of functions of the $L^d$ variables $\eta_{\Gamma_L}$.  We can apply Lemma~\ref{compactness_lemma_2} to find a decreasing sequence $\delta_i \to 0$ and an increasing sequence $K_j \to \infty$ (by applying the diagonal subsequence trick, we can choose them to be independent of $x$ and $L$) so that
\begin{equation}\label{psiDef}
\psi^x_L (\eta) := \lim_{i \to \infty} \lim_{j \to \infty} \phi^x_{L,K_j,\delta_i}(\eta)
\end{equation}
exists, and by uniformity of convergence
\begin{equation} \label{GL_exists_as_lim}
G_L(\eta_L) := \lim_{i \to \infty} \lim_{j \to \infty} \left( \gf{L,K_j}{\delta_i}(\eta_L) - \Av \gf{L,K_j}{\delta_i} \right)
\end{equation}
also exists with
\begin{equation} \label{psi_is_deriv}
\frac{\partial G_L}{\partial \eta_x} =  \psi^x_L(\eta)
\end{equation}
for all $L$, $\eta$, $x \in \Gamma_L$.  Property~\ref{ct_lip} and the relation of FKG to monotonicity then follow from Equation~\eqref{phiDef}, and property~\ref{ct_lip2} follows from Equation~\eqref{2LBound}.

To show property~\ref{ct_M}, we use Equations~\eqref{psi_is_deriv} and~\eqref{psiDef} to write
\begin{equation}
  \Av G_1'=\Av \psi_1^0 = \Av \lim_{i \to \infty} \lim_{j \to \infty} \phi^0_{1,K_j,\delta_i}.
\end{equation}
Recalling the definition of $\phi$ and using dominated convergence to exchange limits and averages we have
\begin{equation}
  \Av G_1' = \lim_{i \to \infty} \lim_{K \to \infty} \frac{\epsilon}{2} \Av \left[ \state{\lop_0}^{h+\delta_i}_K - \state{\lop_0}^{h-\delta_i}_K \right] 
\end{equation}
and since the random fields are i.i.d.\ we can apply the Birkhoff ergodic theorem~\cite{Birkhoff} to replace the disorder average with a volume average, which is related to the derivative of the free energy by Proposition~\ref{LongRange}, giving
\begin{equation}
 \Av G_1' = \frac{\epsilon}{2} \left( \frac{\partial \F}{\partial h+} - \frac{\partial \F}{\partial h-} \right).
\end{equation}

Note that Equation~\eqref{phiDef} implies that, for any $L'~<~L~\le~K$,
\begin{equation} \label{phiMartin}
\phi^x_{L',K,\delta} = \condAv{\phi^x_{L,K,\delta}}{\eta_{\Gamma_{L'}}},
\end{equation}
and we apply the conditional form of the dominated convergence theorem to the limits used to define $\psi$ to obtain
\begin{equation}\label{psiConditional}
\psi^x_{K} = \condAv{\psi^x_L}{\eta_K}.
\end{equation}

Since $\gf{L,K_j}{\delta_i}$ was defined in terms of periodic boundary conditions and the limits used to obtain $\phi$ are independent of $x$, we also have 
\begin{equation}
\phi^x_{L,K,\delta}( T_y \eta) = \phi^{x-y}_{L,K,\delta}(\eta),
\end{equation}
and taking the limit~\eqref{psiDef} we have
\begin{equation}
\psi^x_{L}( T_y \eta) = \psi^{x-y}_{L}(\eta).
\end{equation}
Together with Equations~\eqref{psiConditional}, this lets us obtain 
\begin{equation}
\condAv{\psi_L^y}{\eta_\Lambda}=\condAv{\psi_L^{y+x} \circ T_x }{\eta_{L'}}
\end{equation}
or in other words (in light of~\eqref{psi_is_deriv})
\begin{equation}
\condAv{\frac{\partial G_L}{\partial \eta_y}}{\eta_\Lambda}=\frac{\partial}{\partial \eta_y}{G_{L'}}(T_x \eta)
\end{equation}
The uniform bounds proven above allow us to take the partial derivative with respect to $\eta_y$ outside the conditional average~\cite{Klenke}, and the resulting expression can then be integrated to prove Property~\ref{ct_covariance}.

\subsubsection{Absolutely continuous random fields}

With $\beta = \infty$, it is still obvious from Equation~\ref{GdeltaDef} that $\gf{L,K}{\delta}(0)=0$; Inequality~\eqref{thetaBound} implies $\lip{\gf{L,K}{\delta}} \le \epsilon$, so we can apply Lemma~\ref{compactness_lemma_1} to obtain~\eqref{GL_exists_as_lim}, with $\lip{ G_L}\le \epsilon$.  $\Av G_L =0$ is obvious.  We can apply the diagonal subsequence trick to obtain sequences independent of $L$, which implies that the consistency condition~\ref{consistency_assumption} of Theorem~\ref{CLT_prop} is satisfied.

The difference arises in proving that $\Av G_1'=\epsilon M$.  Without equicontinuity of the derivatives, there is no reason to expect 
that  functions  constructed  like  $\phi$ of the previous section will converge uniformly, or 
 that the pointwise limit will be  differentiable.  
 This problem can be resolved applying the following convergence criteria:

\begin{lem}\label{weak_conv_derivatives_thm}
Let $g$ be a measurable function and $g_n$ a sequence of measurable functions such that $\|g_n\|_\infty\le 1$, $\|g\|_\infty\le 1$, and
\begin{equation}
\lim_{n \to \infty} \int_a^b g_n(t) dt = \int_a^b g(t) dt
\end{equation}
for all $a,b \in \R$.  Then for any finite measure $\nu$ which is absolutely continuous with respect to the Lebesgue measure,
\begin{equation} \label{weak_conv_conclusion}
\lim_{n \to \infty} \int g_n d\nu = \int g\ d\nu.
\end{equation}
\end{lem}
This can be proven by elementary measure theory techniques\cite{RLG.thesis}.

Thanks to Rademacher's theorem~\cite{Heinonen}, Lipschitz continuity guarantees that $f_n$ have distributional derivatives, i.e.\ functions $g_n$ satisfying
\begin{equation}
\int_a^b g_n(t) dt = f_n(b)-f_n(a)
\end{equation}
for any $a,b \in \R$; and furthermore that $\|g_n\|_\infty = \lip{f_n} \le 1$.  We therefore have
\begin{cor}\label{weak_conv_corr}
Let $f_n$ be a sequence of functions $\R \to \R$ such that $f_n \to f$ pointwise, and $\lip{f_n} \le 1$.  Then $f$ has a distributional derivative $f'$ satisfying
\begin{equation}
\lim_{n \to \infty} \int g_n(t) d\nu(t) = \int f'(t) d\nu(t).
\end{equation}
\end{cor}

Applying Corollary~\ref{weak_conv_corr} twice to $\gf{1,K_j}{\delta_i}(\eta_L) - \Av \gf{1,K_j}{\delta_i}$ and $G_1$, we obtain
\begin{equation}
\Av G_1' = \lim_{i \to \infty} \lim_{j \to \infty} \Av \left( \gf{1,K_j}{\delta_i}\right)'(\eta)
\end{equation}
and applying Proposition~\ref{LongRange} and the Birkhoff ergodic theorem to the right hand side as in the finite temperature case we obtain
\begin{equation}
\Av G_1' =  \frac{\epsilon}{2} \left( \frac{\partial \F}{\partial h +} - \frac{\partial \F }{\partial h -} \right)
\end{equation}
and the proof of Theorem~\ref{th:convergence} is complete.

\subsection{Concluding the proofs of Theorems~\ref{alt_main_prop} and~\ref{alt_continuous_prop}}


In summary, we have established that under the hypotheses of Theorems~\ref{alt_main_prop} and~\ref{alt_continuous_prop} the free energy fluctuations $G_L(\eta)$ are uniformly bounded above by $C L^{d-2}$, and correspondingly $C L^{d/2}$, and that if there is a $1^{st}$ order phase transition (i.e.\ a discontinuity in $\frac{\partial \F}{\partial h}$) then $G_L(\eta)$ has asymptotically normal fluctuations on a scale of $L^{d/2}$.  That is is a contradiction even in the marginal dimensions (where the powers match) as a normally distributed random variable will take  with positive probability values which are arbitrarily large on any given scale. 
An alternative  way to present this case is to note that assuming the existence of a $1^{st}$ order phase transition, by Proposition~\ref{CLT_prop} we get 
\begin{equation}\label{mgf_CLT}
\lim_{L \to \infty} \Av \exp \left( t G_L / L^{d/2} \right) = \exp( t^2 b^2 / 2) \, .
\end{equation}
At the same time, if  $\|G_L\|_\infty \le A L^{d/2}$ then for all positive $t$: 
\begin{equation}\label{mgf_bound}
\Av e^{t G_L/L^{d/2}} \le e^{tA}  \, .
\end{equation}
Clearly if $b \ne 0$, the two relations are incompatible at sufficiently large $t$.  Finally  Proposition~\ref{AW_theta} allows to conclude that in  any of the cases under consideration, $b=0$ implies $M=0$, and the proof is complete.

\appendix
\renewcommand{\theequation}{A.\arabic{equation}}
\section{Decay conditions for long range interactions 
}\label{LR_appendix}

In this Appendix we turn to the general class of interactions under which 
the results derived in this work apply.   

The estimate which is relevant for Theorem~ \ref{alt_main_prop} is valid if: 
%
\begin{equation} \label{eq:decay}
\sum_{\substack{ A \ni 0 \\ \diam A \le L }} \diam A \frac{|\partial A|}{|A|} \|Q_A\| \le c' L^{(2-d)/2}
\end{equation}
for some constant $c'$.   In case of two body interaction, that is when $Q_A$ is non zero only when $A$ is a two point set, this condition is met  if 
\be  \label{eq:3d/2}
\|Q_{\{x,y\}}\| \  \le \  \textup{Const.} / |x-y|^{3d/2} \,.  
\ee 

As noted in the proof, the essential estimate which is required for Lemma \ref{simple_UBound_lemma}, and therefore Theorem \ref{alt_main_prop}, 
is that 
\begin{equation}\label{short_range_assumption}
\Av \left\| V_{\Lambda_L}^{\eta} \right\| \le c L^{d-1} + b L^{d/2} \, , 
\end{equation}
with some finite constants $b$ and $c$. 
The contribution in  $ V_{\Lambda_L}^{\eta}$ of terms involving the random fields $\eta$   consists of a collection of $O(L^{d-1})$ independent summands with finite average norm.  Thus the assumption~\eqref{short_range_assumption} requires only a bound of the same form on the nonrandom part of the interaction.  
The following simple calculation~\cite{AW.CMP} 
shows that \eqref{eq:decay} provides a sufficient condition for that. 

By the triangle inequality
\begin{equation}
\|V_L^{0,0}\| \le \sum_{A: \overlappingClass{A}{\Gamma_L}}  \left\| Q_A \right\| \, .  
\end{equation}
In this sum the terms with diameter $L$ or less contribute at most
\begin{equation}
\sum_{\substack{ A \ni 0 \\ \diam A \le L}} 2 d L^{d-1} \frac{\diam A}{|A|} \le 2 d c' L^{d/2} \, ,
\end{equation}
and the remaining portion is bounded by
\begin{equation}
\begin{split}
\sum_{\substack{A \ni 0 \\ \diam A \le L}} & L^d \frac{|\partial A |}{|A|} \| Q_A \|
\\ & \le L^{d-1} \sum_{\substack{A \ni 0 \\ \diam A \le L}} \diam A \frac{|\partial A|}{|A|} \| Q_A \| \le c' L^{d/2} \, . 
\end{split}
\end{equation}
Putting the two parts back together we have Inequality~\eqref{short_range_assumption} with $C' = (2 d + 1) c'$.

The recent result of  Cassandro, Orlandi and Picco~\cite{Cassandro} allows to conclude that at $d=1$ the condition \eqref{eq:3d/2} does indeed provide the threshold decay rate for the validity of 
Theorem~\ref{alt_main_prop}.   They prove  that the phase transition is stable under weak disorder in a family of one dimensional Ising spin systems with long range interactions with decay rates  arbitrarily close to $3/2$, more explicitly with 
$\|Q_{\{x,y\}}\| \  \approx \  \textup{Const.} / |x-y|^{3/2 - \gamma} $  at any $\gamma \in (0,0.08) $.  (Curiously, while the proof of the complimentary result does not extent beyond this narrow range of values of $\gamma$, the lower end of the interval coincides with \eqref{eq:3d/2}.)  The question whether the above criterion is optimal also for $d=2$ is of interest, as there are some notable systems with inverse cube interactions in two dimensions~\cite{SchSt.JPCM}.\\ 


For the stronger conclusion which is derived  here for interactions with continuous symmetries the relevant assumption is not \eqref{short_range_assumption}
 but   
 \begin{equation}\label{continuous_short_range_assum}
\sum_{A \ni 0} (\diam A)^2 |A| \|Q_A\| < \infty  \,    ,  
\end{equation}
which is used in~\eqref{continuous_interaction_bound}, in the proof of the  free energy estimate of Lemma~\ref{th:continuous_upper_bound}.   
For pair interactions this reduces to the statement
\begin{equation}
\sum_{x \in \Zd} \|Q_{\{0,x\}}\| \: |x|^2 < \infty \, , 
\end{equation}
found (in slightly different notation) in Ref.~\onlinecite{QIMLetter}.


\begin{acknowledgments}
We thank S. Chakravarty, R. Fisch, and M. Schechter for discussions; M.A.\ also wishes to thank the Departments of Physics and Mathematics at the Weizmann Institute of Science for  hospitality.
The work was supported in part by NSF Grants DMS-0602360 (M.A.) and DMR 08-02120 (R.L.G.\ and J.L.L.), ERC Starting Grant CoMBoS-239694 (R.L.G.), and by AFOSR Grant FA9550-10-1-0131 (R.L.G.\ and J.L.L.).

\end{acknowledgments}

%


%

\end{document}